\documentclass[twocolumn]{aastex631}

\usepackage{amsmath}
\usepackage{soul}
\usepackage{multirow}

\definecolor{mygreen}{RGB}{0,153,76}

\newcommand*\df{\mathop{}\!\mathrm{d}}
\newcommand{\n}{\nonumber}

\begin{document}

\title{Transient gamma rays from the 2021 outburst of the recurrent nova RS Ophiuchi:\\ the effect of gamma-ray absorption}


\author[0000-0002-5611-095X]{Vo Hong Minh Phan}
\affiliation{Sorbonne Universit\'e, Observatoire de Paris, PSL Research University, LUX, CNRS, 75005 Paris, France}

\author[0000-0002-5856-7662]{Pierre Cristofari}
\affiliation{LUX, Observatoire de Paris, Université PSL, Sorbonne Université, CNRS, 92190 Meudon, France}

\author[0000-0003-0543-0467]{Enrico Peretti}
\affiliation{INAF - Astrophysical Observatory of Arcetri, Largo E. Fermi 5, 50125 Florence, Italy}
\affiliation{APC, Universit\'e Paris Diderot, CNRS/IN2P3, CEA/Irfu, Observatoire de Paris, Sorbonne Paris Cit\'e, France}

\author{Vincent Tatischeff}
\affiliation{Universit\'{e} Paris-Saclay, CNRS/IN2P3, IJCLab, 91405 Orsay, France}

\author[0000-0001-5865-6772]{Andrea Ciardi}
\affiliation{Sorbonne Universit\'e, Observatoire de Paris, PSL Research University, LUX, CNRS, 75005 Paris, France}




\begin{abstract}
In 2021, RS Ophiuchi was the first nova to be detected in the very-high-energy (TeV) gamma-ray domain, directly testifying of efficient acceleration of charged particles up to at least the TeV range at the nova shock. Surprisingly, the TeV gamma-ray signal peaks $\sim 2$ days after the GeV signal and the origin of this delay has still not been clearly understood. We investigate the possibility that this delay is due to the effect of gamma-ray absorption resulted from interactions between gamma rays and optical photons copiously emitted during the outburst. We model particle acceleration at a nova shock to obtain the gamma-ray emission produced in interactions between the accelerated particles and the shocked gas. The effect of gamma-ray absorption is then included in details using the radiative transfer equation. We find that this can naturally account for the delay between the peaks of GeV and TeV gamma-ray lightcurves. This result emphasizes the importance of gamma-ray absorption for interpreting gamma-ray observations of novae in the TeV range which, in turn, demonstrates the necessity of a multi-wavelength view for unraveling the underlying physics of particle acceleration in these systems.  
\end{abstract}

\keywords{Recurrent novae (1326), Gamma-ray transient sources (1853), Shocks (2086)}


\section{Introduction}
Novae are transients observed in many different wavelengths from radio to X-rays and gamma rays. In classical and recurrent novae, the increased brightness results from thermonuclear explosions on the surface of white dwarfs (WDs) embedded in stellar winds. These explosions lead to formation of shocks, which can accelerate particles up to TeV energies or beyond in certain cases \citep{metzger2015,metzger2016}. The presence of these high-energy particles indeed means that non-thermal emissions, especially gamma rays induced by leptonic processes (e.g. bremsstrahlung or inverse Compton scattering) and hadronic processes (e.g. decay of neutral pions produced in proton-proton interactions) are expected from these sources. 

Many novae were, in fact, observed in GeV gamma-rays by the Fermi-LAT gamma-ray telescope \citep[see e.g.,][for a review]{chomiuk2021}. Recently, both HESS and MAGIC gamma-ray telescopes reported the detection of TeV gamma rays from the 2021 outburst of the recurrent nova RS Ophiuchi (RS Oph) making this system the first nova ever to be observed in the TeV gamma-ray domain \citep{aharonian2022,acciari2022}. RS Oph consists of a WD embedded in the wind of a red giant (RG). The WD continuously accretes materials from the RG wind and a nova explosion occurs when the outer layer of the WD reaches a critical state. In the case of RS Oph, the nova explosions occur approximately every 10 or 20 years \citep{schaefer2010}. The most recent explosion was in 2021 with the shock speed inferred by X-ray and radio observations to be around a few thousands km/s \citep{cheung2022,pandey2022}, which is of the same order of magnitude as in the previous explosion in 2006 \citep{das2006,bode2006,sokoloski2006}. The relatively high shock speed, the strong magnetic field strength (believed to be of order of a few Gauss close to the RG), and their detection in gamma rays from GeV to TeV energy, make novae ideal sources for the study of particle acceleration (see e.g. \citealt{martin2013,martin2018}). 

Interestingly, gamma-ray observations revealed a time-delay between the peaks of the GeV and TeV lightcurves.
It has been suggested in some studies that the standard acceleration model, where the injection spectrum of accelerated particles around the shock follows a simple power law in momentum with an exponential cut-off around the maximum particle energy, cannot explain this feature and, thus, such a time-delay can potentially provide new insights into the process of shock acceleration. In fact, many of the previous works have attempted to explain this feature using a more gradual exponential cut-off for the shock injection spectrum \citep{aharonian2022,zheng2022}. In other approaches,  the typical shock injection spectrum is kept but multiple populations of accelerated particles have to be introduced. For example, \cite{diesing2023} suggest the presence of two shocks where the slower one dominates the GeV gamma-ray emissions in the first one or two days of the outburst and the faster shock accounts for the TeV emissions at later time. \citet{desarkar2023}, on the other hand, put forward a lepto-hadronic scenario where the dominant contribution of GeV and TeV emissions are induced respectively by electrons and protons. 

Since these models have to introduce more parameters than a typical shock acceleration model in order to obtain good fits for the delay between the peaks of the GeV and TeV lightcurves, we would like to offer an alternative explanation, which relies on the effect of gamma-ray absorption. Given that the nova is also very bright in optical light, TeV gamma rays can be absorbed as they interact with optical photons. In fact, the role of gamma-ray absorption has previously been studied most notably by \citet{acciari2022} and \citet{diesing2023} where the authors concluded that this effect should be negligible based on their approximate estimates of the gamma-ray opacity. However, none of the above-mentioned works have performed a thorough analysis with the radiative transfer of gamma rays taken into account to verify if modifications of the standard particle acceleration model are actually required. The aim of this work is to fill in this gap.

We study gamma-ray emission from the 2021 outburst of RS Oph using a model of particle acceleration in a single nova shock. We choose $t=0$ day at one day before the optical peak of the 2021 outburst as in \cite{aharonian2022}, which corresponds to Modified Julian Day 59434.25 (or 2021 August 8.25 in Coordinated Universal Time). The paper is organized as follows. We first discuss the dynamics of nova shock and the corresponding acceleration process of non-thermal particles. The gamma-ray emission from accelerated hadrons interacting with the compressed material of the RG wind is then modeled. Finally, we include the gamma-ray absorption due to interactions between gamma rays and optical photons, using the radiative transfer equation. We find that the gamma-ray absorption can naturally account for the time-delay of $\sim$ 2 days between the GeV and TeV lightcurves.  

\section{Particle acceleration in nova shocks}
\subsection{Dynamics of nova shocks}
The gas mass density profile at a distance $r$ from the WD can be modeled as follows \citep{aharonian2022}
\begin{eqnarray}
    \rho(r)\simeq\frac{\dot{M}}{4\pi v_{\rm wind} \left(r^2_{\rm orb}+r^2\right)}, \label{eq:rho}
\end{eqnarray}
where $\dot{M}$ is the mass-loss rate of the RG, $v_{\rm wind}$ is the wind speed, and $r_{\rm orb}$ is the orbital radius of the RG. Recent analyses of H$\alpha$ emission line and Na I D absorption lines indicate a mass-loss rate of about $5\times 10^{-7}\,M_{\odot}/{\rm yr}$ \citep{booth2016} which, for the typical wind speed for the RG between $10$ km/s and $30$ km/s \citep{lamers1999}, gives the ratio $\dot{M}/v_{\rm wind}$ of about a few times $10^{13}$ g/cm. This ratio determines the normalization of the gamma-flux as we shall see later, but it is degenerate with several other parameters. 
Thus, for simplicity, we will fix this ratio to $\dot{M}/v_{\rm wind}=2\times 10^{13}$ g/cm for the fit of gamma-ray data similar to the value used in modeling multi-wavelength data in the previous outburst by \citet{tatischeff2007}. 
As for the orbital radius, it is relatively well constrained to the value of about $1.48$ au with spectroscopic optical observations \citep{brandi2009}.         

The nova shock speed $v_{\rm sh}$ is expected to evolve as follows  
\begin{eqnarray}
    v_{\rm sh}(t)=\left\{ \begin{array}{ll} 
    v_{\rm sh,0} & t\leq t_{\rm r}\, , \\
    \displaystyle v_{\rm sh,0}\left(\frac{t}{t_{\rm r}}\right)^{-\alpha} & t>t_{\rm r}~, \end{array} \right.
    \label{eq:vsh}
\end{eqnarray}
where initially the shock expands with a constant speed in the free expansion phase and then, at $t\simeq t_{\rm r}$, it transitions into the radiative phase where $v_{\rm sh}\sim t^{-\alpha}$ with $\alpha\sim 0.5$. It has been suggested also that there might exist a Sedov-Taylor phase (or adiabatic phase) where $v_{\rm sh}\sim t^{-1/3}$ \citep{bode1985,das2006}. Optical and X-ray observations, however, indicate that this phase may be too brief to be detected. We note also that, in the above equations for shock evolution, the eruption time has been implicitly assumed to happen at $t_{\rm er}=0$ day. We checked, however, that the fit to gamma-ray data is not very sensitive to the exact value of $t_{\rm er}$ as long as $t_{\rm er}\lesssim 0.2$ day. 

Shock speeds inferred from post-shock gas temperature measurements using X-ray observations seem to indicate $v_{\rm sh,0}\simeq 2470$ km/s, $\alpha\simeq0.43$ and  $t_{\rm r}\simeq 6$ day (\citealt{cheung2022}, see also \citealt{orio2023}). Spectroscopic data from optical observations, on the other hand, are well fitted for $v_{\rm sh,0}\gtrsim 4000$ km/s, $\alpha\simeq 0.6$ and $t_{\rm r}\simeq 4$ day \citep{pandey2022}. In fact, \cite{tatischeff2007} have pointed out that back-reactions of particles accelerated in nova shocks can modify relations between shock speeds and post-shock gas temperatures and explain the differences between shock properties inferred by X-ray and optical observations. In the following, we will treat $v_0$, $t_{\rm r}$, and $\alpha$ as free parameters that can be obtained by fitting simultaneously the shock speed profile inferred from optical and X-ray observations and the gamma-ray emissions.       
 
\subsection{Maximum proton energy}
The acceleration rate of protons in shocks can be estimated as \citep[see e.g.][]{tatischeff2007}
\begin{eqnarray}
    \dot{E}_{\rm acc}(t)\simeq\frac{qB_2(t)v^2_{\rm sh}(t)}{\eta c},
\end{eqnarray}
where $q$ is the charge of proton, $B_2$ is the magnetic field strength downstream of the shock, $c$ is the speed of light, and $1/\eta$ is the acceleration efficiency. It is generally suggested that $\eta\gtrsim 2\pi$ but the exact value is not well known. Since $\eta$ is degenerate with $B_2$, we can fix $\eta=10 \pi$ for simplicity (roughly similar to the acceleration efficiency assumed by \citealt{aharonian2022}). We will assume $B_2$ to be the compressed background magnetic field around the RG as in \cite{aharonian2022} with the following profile
\begin{eqnarray}
    B_2= B_{2,0} \left(\frac{\sqrt{r_{\rm orb}^2+R_{\rm sh}^2}}{0.35\,{\rm au}}\right)^{-2},\label{eq:Bfield}
\end{eqnarray}
where $B_{2,0}$ ranging from 1 to 10~G and $R_{\rm sh}(t)=\int_0^t v_{\rm sh}(t')\df t'$ is the shock radius. 

Particles are also losing energy over time due to adiabatic expansion of the shock. The energy loss rate can be modeled as follows \citep{caprioli2010,cristofari2020}
\begin{eqnarray}
    \dot{E}_{\rm ad}(E,t)\simeq \frac{pv}{\left(\rho v_{\rm sh}^2(t)\right)^{\frac{1}{3\gamma_{\rm gas}}}}\frac{\df}{\df t}\left[\left(\rho v_{\rm sh}^2(t)\right)^{\frac{1}{3\gamma_{\rm gas}}}\right],
\end{eqnarray}
where $\gamma_{\rm gas }=5/3$ is the adiabatic index of the gas. 

The maximum kinetic energy of protons at any time $t$ can then be evaluated by solving
\begin{eqnarray}
    \frac{\df E_{\rm max}}{\df t}=\dot{E}_{\rm acc}(t)+\dot{E}_{\rm ad}(E_{\rm max},t).\label{eq:Emax}
\end{eqnarray}
Once the nova shock dynamics and the maximum energy of accelerated particles are defined, we can proceed to discuss the spectrum of these particles accumulated around the shock.  

\subsection{Cumulative proton spectrum}
The injection spectrum of protons around the shocks is typically assumed to be a power law in momentum $p$ with an exponential cut-off around $p_{\rm max}$ which can be estimated from $E_{\rm max}$ with Eq. \ref{eq:Emax}. The spectrum is also normalized such that, at any given time, the pressure of accelerated protons is a fraction $\xi_{\rm p}$ of the shock ram pressure \citep{cristofari2021}, which gives 
\begin{eqnarray}
    f_{\rm p}(E,t)=\frac{3\xi_{\rm p}\rho v_{\rm sh}^2 }{m_{\rm p}^2c^4 \beta I(\delta)}\left(\frac{p}{m_{\rm p}c}\right)^{2-\delta}\exp\left(-\frac{p}{p_{\rm max}}\right).
\end{eqnarray}
Here, $m_{\rm p}$ is the proton mass, $\rho$ and $v_{\rm sh}$ are estimated at the shock radius, and $I(\delta)=\int^{x_{\rm max}}_{x_{\rm min}} \df x x^{4-\delta}{\rm e}^{-x/x_{\rm max}}/\sqrt{1+x^2}$ with $x=p/(m_{\rm p} c)$. The cumulative spectrum of protons accelerated at the nova shock can then be obtained by solving the following transport equation \citep[see e.g.][]{aharonian2022}:
\begin{eqnarray}
    \dfrac{\partial N_{\rm p}(E,t)}{\partial t} + \dfrac{\partial }{\partial E}\left[\dot{E}_{\rm ad} (E,t) N_{\rm p}(E,t)\right]\n\\
    \qquad\qquad\qquad = \dfrac{4\pi R^2_{\rm sh}(t) v_{\rm sh}(t)}{r_{\rm c}} f_{\rm p}(E,t),
\end{eqnarray}
where $r_{\rm c}$ is the shock compression ratio. We will assume $r_{\rm c}\simeq 4$ as commonly adopted for strong shocks \citep{cristofari2021}.

\section{Gamma-ray emissions from nova shocks}

\begin{figure}[h!]
    \centering
    \includegraphics[width=\columnwidth]{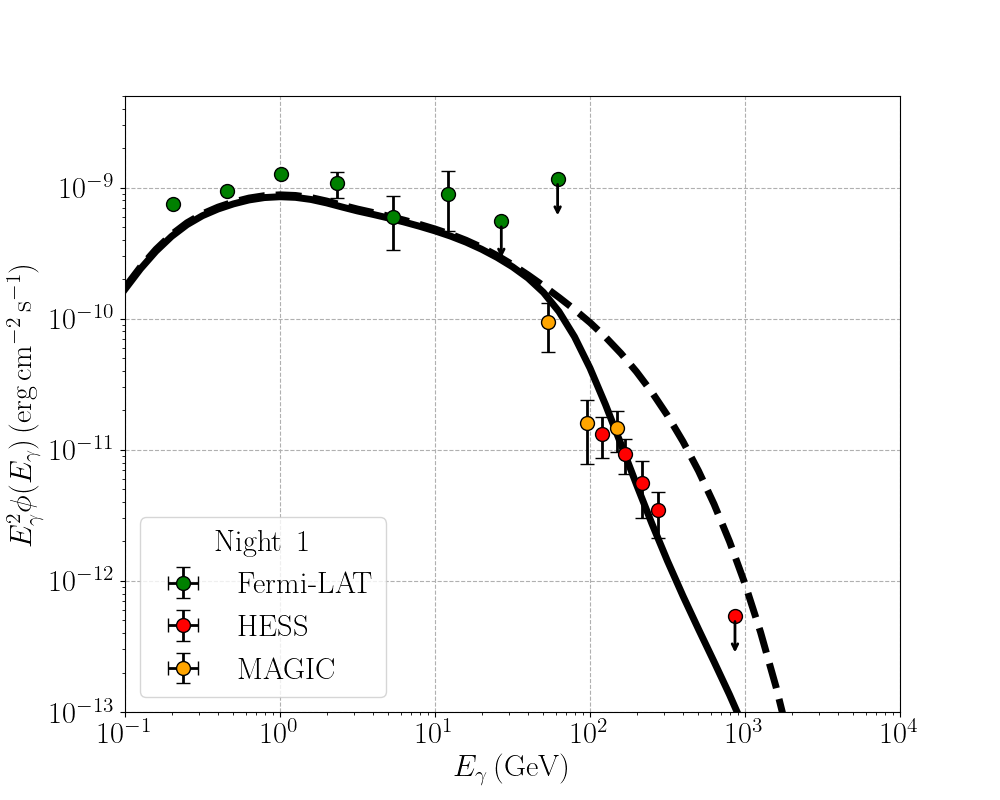}
    \includegraphics[width=\columnwidth]{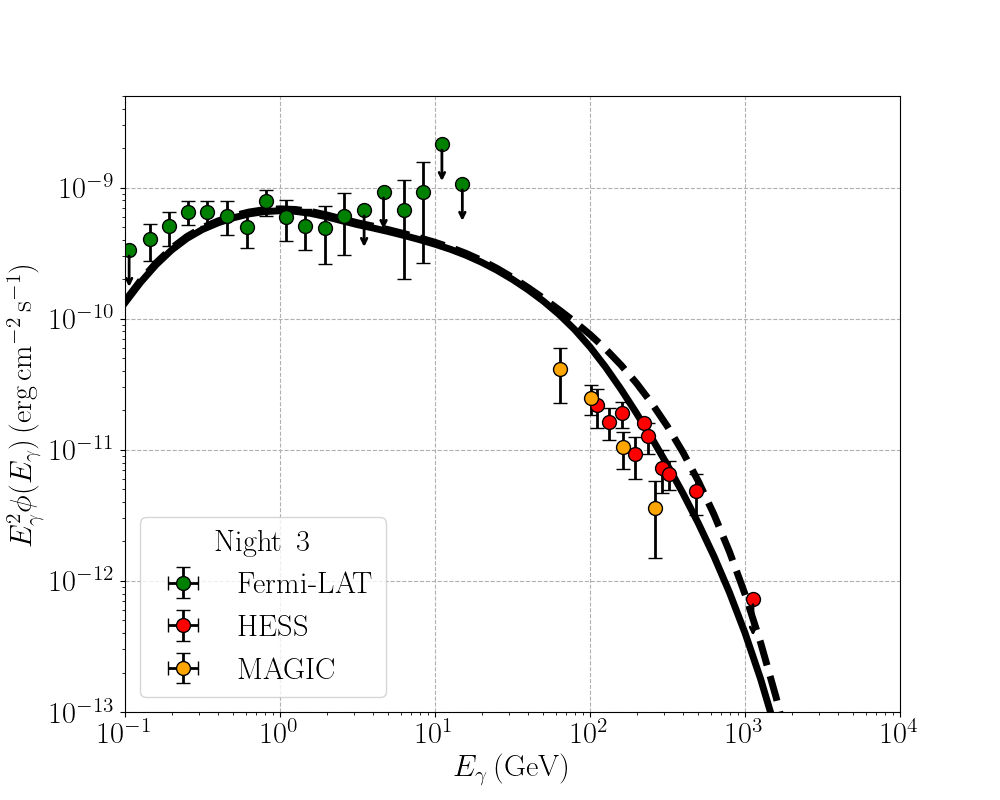}
	\includegraphics[width=\columnwidth]{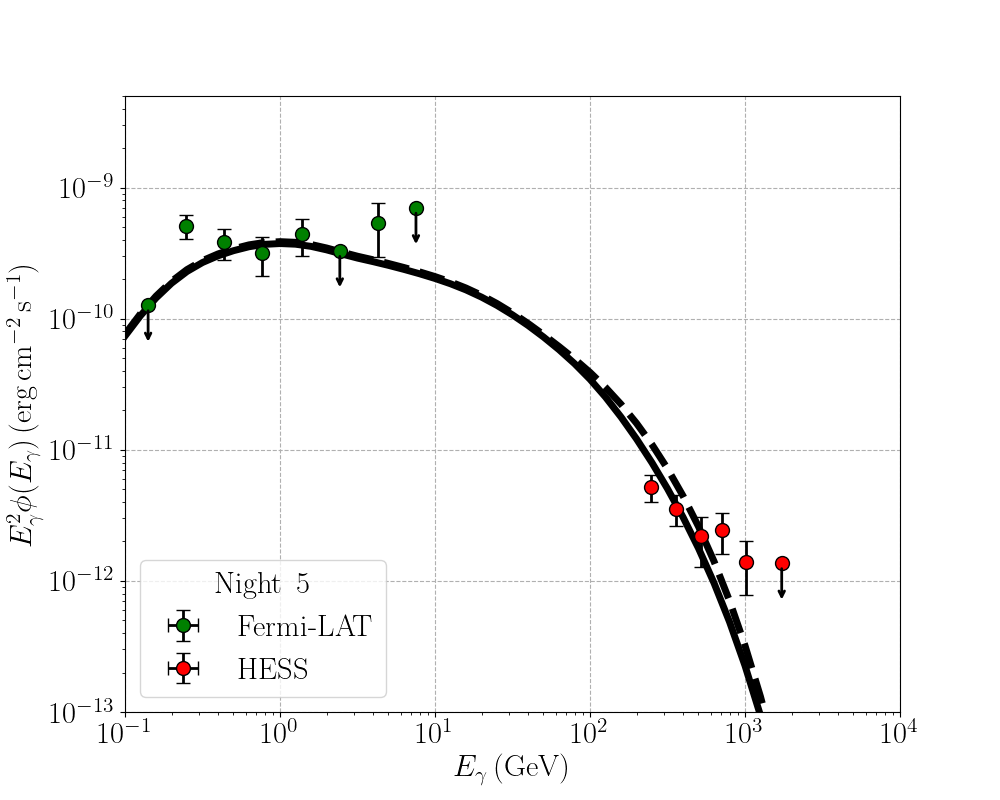}
    \caption{Gamma-ray spectra compared to data from \cite{aharonian2022, acciari2022}. Solid and dashed lines are for the case with and without gamma-ray absorption respectively.}
    \label{fg:gamma}
\end{figure}

\subsection{Gamma-ray production}
Interactions between accelerated protons and ambient protons (materials swept up by the nova shock) can lead to production of neutral pions which decay quickly into gamma rays. These gamma rays can be partially absorbed by interactions with optical photons produced from the outburst (see the next subsection). However, previous works \citep[e.g.][]{diesing2023}, based on their approximate estimates of the gamma-ray opacities, suggest that gamma-ray absorption is negligible such that the expected gamma-ray flux from the nova can be estimated as follows
\begin{eqnarray}
    \phi_0(E_\gamma,t)\simeq\frac{\rho_{\rm sh}(t)}{4\pi d_{\rm N}^2 m_{\rm p}}\int^{E_{\rm max}}_{E_{\rm min}}\df E N_{\rm p}(E,t)\n\\
    \qquad\qquad\times v_{\rm p}\varepsilon_n(E)\frac{\df \sigma_{\rm pp}(E,E_\gamma)}{\df E_\gamma}, 
    \label{eq:phi0}
\end{eqnarray}
where $\rho_{\rm sh}(t)\simeq 4\rho(r=R_{\rm sh}(t))$ is the mass density of shocked gas, $d_{\rm N}$ is the distance of the nova to Earth, $\varepsilon_n(E)$ is the nuclear enhancement factor to take into account gamma rays from accelerated nuclei heavier than proton, and $\df \sigma_{\rm pp}(E,E_\gamma)/\df E_\gamma$ is the differential cross-section for gamma-ray production in proton-proton interaction \citep{kafexhiu2014}.

\subsection{Gamma-ray absorption}
\subsubsection{Distribution of optical photons}
\label{sec:optical}
As mentioned above, since nova outbursts can be very bright in optical light, the medium around these novae may be opaque to gamma rays of energy above about 200 GeV in the first days of the outbursts. Optical photons from the outburst can be modeled with a black-body distribution which can be described with the temperature $T_{\rm opt}$ and the energy density $u(r,t)$. If we assume that optical photons are homogeneously and isotropically emitted within a photosphere of radius $R_{\rm ph}$, it can be shown that the energy density $u(r,t)$ of optical photons follows 
\begin{eqnarray}
    u(r,t)=\frac{L_{\rm opt}(t)}{4\pi c}\frac{3r}{2R^3_{\rm ph}}g_{\rm opt}\left(\frac{r}{R_{\rm ph}}\right),
    \label{eq:uOPT}
\end{eqnarray}    
with
\begin{eqnarray}
    g_{\rm opt}(x)=\frac{1}{x}+\left(\frac{1}{x^2}-1\right)\ln\left(\sqrt{\left|\frac{x+1}{x-1}\right|}\right). \label{eq:gopt}
\end{eqnarray}
Note that we can verify from the above equation that $u(r,t)\sim L_{\rm opt}(t)/(4\pi r^2 c)$ in the limit where $r\gg R_{\rm ph}$ which is commonly adopted for rough estimates of the gamma-ray absorption in previous works \citep[see e.g.][]{diesing2023}.  

For RS Oph 2021, we can adopt $T_{\rm opt}\simeq 1.1\times 10^4$ K \citep{cheung2022}, $R_{\rm ph}\simeq 200 R_{\odot}$ \citep{acciari2022}, and the optical lightcurve $L_{\rm opt}(t)$ can be inferred from the observed optical spectra. In fact, \citet{cheung2022} have already performed the analyses for the optical spectra taking into account the interstellar absorption to derive the optical luminosity over time for the 2021 outburst of RS Oph. 
These analyses assume, however, that the distance from RS Oph to Earth is about $1.6$ kpc. This distance, which was first estimated using HI absorption line \citep{hjellming1986}, is quite commonly adopted in the literature. However, such value means that the expected accretion rate might be much smaller than required for the typical recurrence period of RS Oph \citep{schaefer2009}. More recent estimates of the distance, e.g. using the high-resolution radio observations together with the measurements of the shock speed profile \citep{rupen2008} or the parallax distance provided by Gaia \citep{gaia2021} lead to values around $2.45$ or $2.68$ kpc (see \citealt{acciari2022} for a more complete discussions). 

In the following, we will follow \citet{acciari2022} and choose $d_{\rm N}=2.45$ kpc for the fit of the gamma-ray emission. In order to precisely obtain the optical luminosity over time at $d=2.45$ kpc, we need to re-analyse the optical spectra and correct for interstellar absorption of the source emission at this distance. Note however that RS Oph is at the latitude of $10.37^{\circ}$, meaning that the correction factor for interstellar absorption should be rather similar for $d_{\rm N}=1.6$~kpc and $2.45$~kpc. For this reason, the optical lightcurve can simply be rescaled with distance for $d_{\rm N}>1.6$ kpc. We then choose the optical lightcurve of the following form for $t\gtrsim 1$ day
\begin{equation}
    L_{\rm opt}(t)\simeq 7.8\times 10^{38} \left(\frac{d_{\rm N}}{1.6\, {\rm kpc}}\right)^2 \left(\frac{t}{1\,{\rm day}}\right)^{-1} \, {\rm erg/s},
    \label{eq:Lopt}
\end{equation}
where the normalization and the scaling with time has been fitted to optical luminosity data derived by \citealt{cheung2022} for $d_{\rm N}=1.6$ kpc (see Fig. \ref{fg:luminosity} in Appendix \ref{appendixA}). 

\subsubsection{Radiative transfer of gamma rays}
\label{sec:rad-transfer}
The effect of absorption can be taken into account by solving the radiative transfer equation for gamma rays. We have assumed that gamma rays are mostly produced in a thin layer around the nova shock which allows us to express the gamma-ray flux in the following form
\begin{eqnarray}    
    \phi(E_\gamma,t)=\frac{\phi_0(E_\gamma,t)}{2}\left[e^{-\tau_1(E_\gamma,t)}+e^{-\tau_2(E_\gamma,t)}\right],
    \label{eq:phi}
\end{eqnarray}
where the exact form of the opacities $\tau_1(E_\gamma,t)$ and $\tau_2(E_\gamma,t)$ are derived in more details in Appendix \ref{appendixA}. 

In this representation, we can interpret roughly that half of the gamma rays propagate through the region with $r\leq R_{\rm sh}$ and experience absorption as represented by the average opacity $\tau_1(E_\gamma,t)$. The other half escape directly into the region $r\geq R_{\rm sh}$ and are attenuated by with the average opacity $\tau_2(E_\gamma,t)$. In fact, $\tau_2(E_\gamma, t)$ is similar to the rough estimate of the gamma-ray opacity provided in many of the previous works \citep[e.g.][]{diesing2023}. However, the exact value of $\tau_2(E_\gamma, t)$ can vary by a factor of a few depending on the chosen value of $d_{\rm N}$ (because $L_{\rm opt}\sim d^2_{\rm N}$, see Eq. \ref{eq:Lopt}) which can be significant for the gamma-ray lightcurve as the effect of absorption is exponential. More importantly, when the full radiative transfer is taken into account, we can see that half of the gamma rays have to pass through the region $r\leq R_{\rm sh}$ and are more strongly absorbed ($\tau_1(E_\gamma,t)>\tau_2(E_\gamma,t)$) since optical photon density is higher close to the WD.

\subsection{Results for the 2021 outburst of RS Ophiuchi}
\label{sec:results}
In this subsection, we present the results of our modeling of the transient gamma-ray emission from a nova with gamma-ray absorption applied for the 2021 outburst of RS Oph. All the parameters of the model are summarized in Table \ref{table:1} in Appendix \ref{appendixB}. Some parameters are fixed to values obtained in previous works since they are already well constrained with other observations. 

In Fig. \ref{fg:gamma}, we present the gamma-ray spectra assuming parameters that best fit the Fermi-LAT, HESS and MAGIC data \citep{aharonian2022, acciari2022}. We found that the effect of absorption is most significant between $t=1$ day and $t=2$ day for gamma rays of energy $\gtrsim 200$ GeV. The gamma-ray spectrum $\gtrsim 200$ GeV at $t\simeq 1.6$ day (corresponding to night 1 in HESS convention) is significantly suppressed such that it matches well with data without the need for modifications of the injection spectrum to suppress the amount of accelerated TeV protons as commonly adopted in previous works. Since proton-proton interactions produce also neutrinos, a potentially observational consequence of our model would be that one might be able to detect TeV neutrinos from novae like RS Oph in the first days of their outbursts even though TeV gamma rays might be undetectable in the same period. However, the predicted flux of TeV neutrinos from RS Oph would still be too low to be detected by current neutrino observatories, such that this phenomenon might be tested only with future instruments.
\begin{figure}
    \centering
	\includegraphics[width=3.6in]{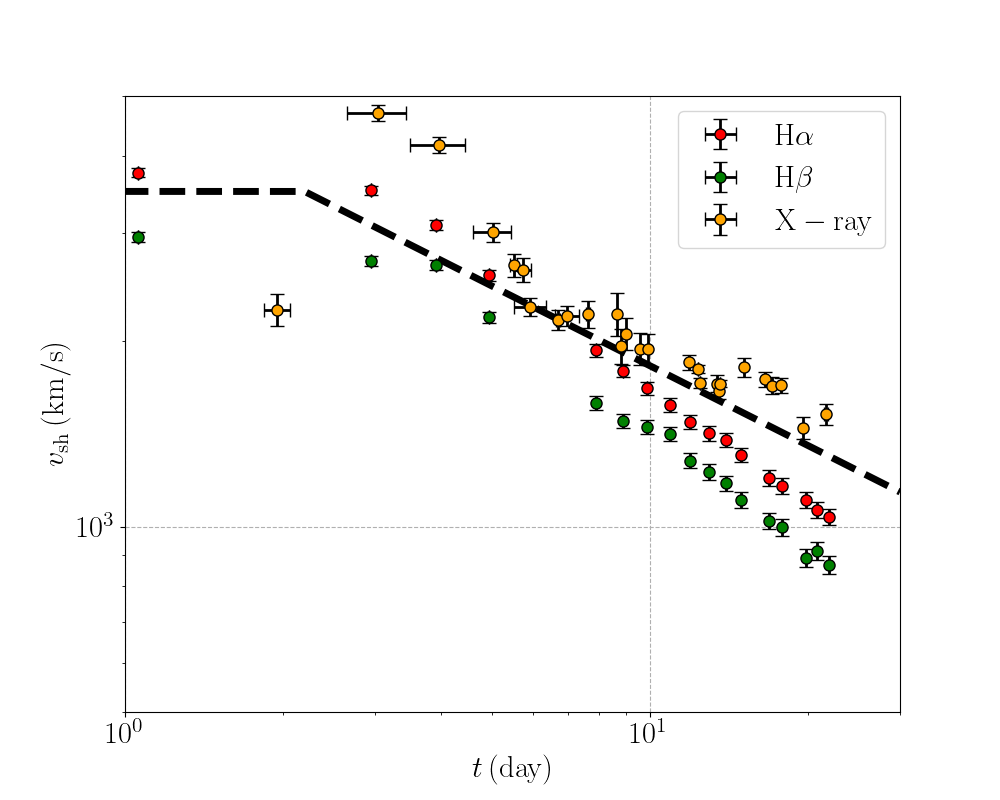}
    \caption{Shock speed evolution in time (dashed line) compared to data inferred from H$\alpha$ and H$\beta$ emission lines (red and green filled circles) by \citet{pandey2022} and from X-ray data (yellow filled circles) by \citet{orio2023}.}
    \label{fg:vsh}
\end{figure}

It is worth mentioning again that the shock speed can also be measured using infrared and X-ray observations (with rather large uncertainty). The shock speed measurements derived from the H$\alpha$ and H$\beta$ emission line profile modeling \citep{pandey2022} and from the post-shock temperature\footnote{We convert the postshock temperature from Fig. 6 of \citet{orio2023} into shock speed using Eq. 3 of \citet{bode2006}.} inferred from observations with NICER X-ray telescope \citep{orio2023} are collected in Fig. \ref{fg:vsh}. The shock speed evolution that best fit both these measurements and the gamma-ray data from Fermi-LAT and HESS are also presented. 

\begin{figure}
    \centering
	\includegraphics[width=3.6in]{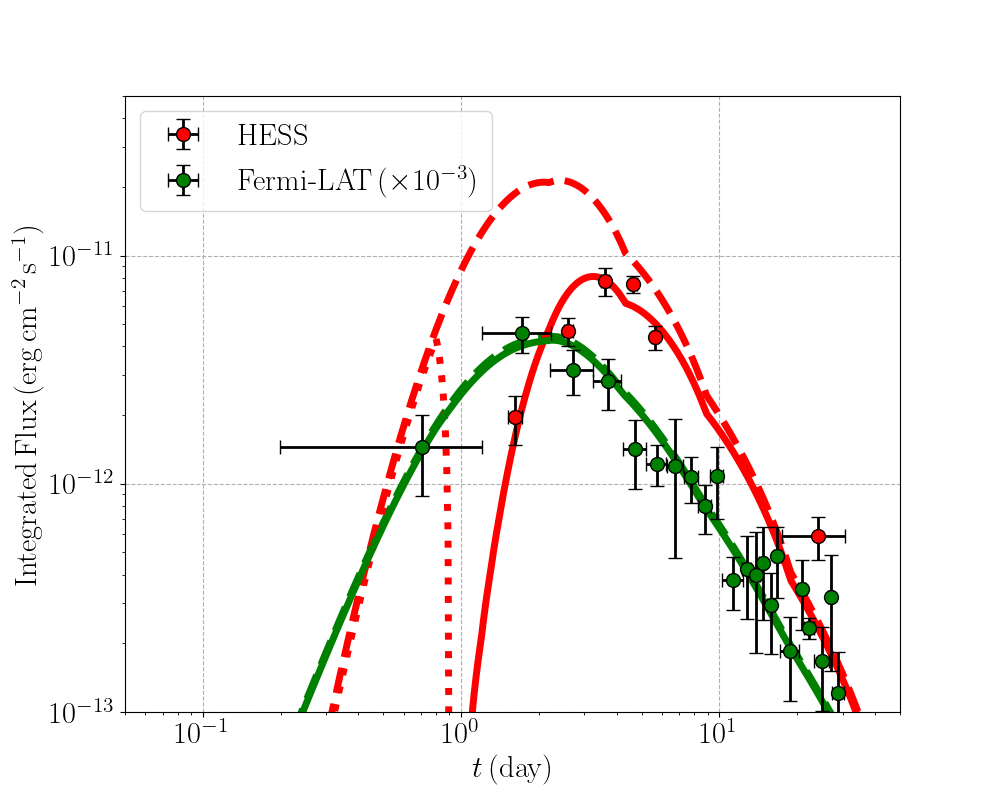}
    \caption{Integrated gamma-ray flux over time compared to data from \cite{aharonian2022}. The lightcurves for Fermi-LAT and HESS (green and red line respectively) are estimated by integrating the gamma-ray flux over the energy ranges 0.1~--~100~GeV and 250~--~2500 GeV, respectively. Solid and dashed lines are correspondingly for the case with and without gamma-ray absorption. The TeV lightcurve with absorption for $t<1$ day is plotted in dotted line to highlight the uncertainty of the model in this period of time where no data is available (see discussion in Section \ref{sec:results}).}
    \label{fg:time_gamma}
\end{figure}

We show also the integrated gamma-ray flux over time in Fig. \ref{fg:time_gamma}. The gamma-ray spectra were integrated in the ranges 0.1~--~100~GeV and 250~--~2500~GeV to obtain lightcurves that can be compared with the GeV and TeV lightcurves measured with Fermi-LAT and HESS. The GeV and TeV lightcurves are presented without  (dashed line) and with (solid line) gamma-ray absorption. In the case without absorption, the predicted GeV and TeV lightcurves peak at the same time. This is because the maximum energy of accelerated particles is reached roughly after $t\simeq 1$ day, which makes the spectral shape of accelerated particles rather similar for $t\gtrsim 1$ day. The lightcurves in both GeV and TeV energy ranges are, therefore, determined mostly by the evolution of the shock and have rather similar time dependence at later time (after reaching the peaks) for the case without absorption. It is clear however that we overestimate the emission for TeV gamma rays at early time in this case. 

A better fit of TeV lightcurve is obtained when gamma-ray absorption is correctly taken into account. Fig.~\ref{fg:time_gamma} shows that the gamma-ray absorption is strongest at $t\simeq 1$ day, which is around the maximum of the optical lightcurve. Note also that the TeV lightcurve in this case is plotted differently (dotted red line) for $t\lesssim 1$ day to indicate that the shape of the TeV lightcurve depends strongly on our interpolation of the optical luminosity for $-2 \,{\rm day}\lesssim t\lesssim 1\,{\rm day}$ (data on the optical luminosity of RS Oph derived by \citealt{cheung2022} are not available in this period). We can also see that the effect of gamma-ray absorption naturally leads to a delay of about 1 or 2 days between the peaks of the GeV and TeV lightcurves. 

\section{Summary and prospects}
We study the transient gamma-ray emissions from the most recent outburst of RS Oph in 2021 with a focus on interpreting the delay between the peaks of GeV and TeV lightcurves. Using a radiative transfer model for gamma rays produced in interactions between particles accelerated around the nova shock and the RG wind materials, we have shown gamma rays passing through the shock downstream are strongly absorbed such that half of the gamma rays with $E_\gamma\gtrsim 200$ GeV are almost completely absorbed in the first one or two days of the outburst. The remaining half escaping directly into the upstream region of the shock are absorbed with an opacity denoted as $\tau_2$ (see Eq. \ref{eq:phi}) comparable to the approximate opacity estimated by, e.g., \citet{diesing2023}. This means that the gamma-ray flux, for $t\lesssim 1$ day and $E_\gamma\gtrsim 200$ GeV, with and without absorption should be different by a factor $\sim e^{-\tau_2}/2$ which can be between $6$ to more than $10$ depending on the exact value of the nova distance $d_{\rm N}=1.4$ kpc or $2.45$ kpc.       

More importantly, the effect of gamma-ray absorption naturally account for the time-delay between the peaks of GeV and TeV gamma-ray lightcurves such that the modifications of the shock injection spectrum or the presence mutiple particle populations are no longer necessary for explaining the data \citep[][]{aharonian2022,diesing2023,desarkar2023}. This result highlights the important role of gamma-ray absorption for better understanding of the TeV gamma-ray emission from novae which is also crucial for deciphering the underlying process of acceleration, especially around the maximum energy of shock-accelerated particles. 

\begin{acknowledgments}
VHMP acknowledges support from the Initiative Physique des Infinis (IPI), a research training program of the Idex SUPER at Sorbonne Universit\'e. PC acknowledges support from the PSL Starting Grant GALAPAGOS.
EP acknowledges support from INAF through ``Assegni di ricerca per progetti di ricerca relativi a CTA e precursori''.
\end{acknowledgments}

\vspace{5mm}



\newpage
\appendix

\section{Gamma-ray opacities}
\label{appendixA}
In order to take into account the effect of gamma-ray absorption, we will first find the gamma-ray intensity along each line of sight $I(E_\gamma, \theta, s\simeq d_{\rm N}, t)$ by solving the radiative transfer equation
\begin{eqnarray}
    \frac{\df I(E_\gamma,\theta,s,t)}{\df s}= -I(E_\gamma,\theta,s,t)\eta_{\rm abs}(E_\gamma,r(\theta,s),t)+I_0(E_\gamma,t)\delta(r(\theta, s)-R_{\rm sh})
    \label{eq:rad_transfer}
\end{eqnarray}
where $r(\theta, s)$, $\eta_{\rm abs}(E_\gamma, r, t)$ and $I_0(E_\gamma,t)$ are respectively the distance from the WD, the absorption coefficient and the gamma-ray intensity around the shock which could be estimated as follows 
\begin{eqnarray}
    && r(\theta,s)=\sqrt{s^2-2s\sqrt{R^2_{\rm sh}-d^2\sin^2\theta}+R^2_{\rm sh}},\label{eq:r(theta,s)}\\
    && \eta_{\rm abs}(E_\gamma,r,t)=\int_{0}^{\infty}\df E_{\rm ph} f_{\rm opt}(E_{\rm ph},r,t)\sigma_{\gamma\gamma}(E_\gamma,E_{\rm ph}),\\
    && I_0(E_\gamma,t)=\frac{\phi_0(E_\gamma,t)}{4\pi}\left(\frac{d_{\rm N}}{R_{\rm sh}}\right)^2.
\end{eqnarray}

\begin{figure}[h!]
    \centering
	\includegraphics[width=0.75\columnwidth]{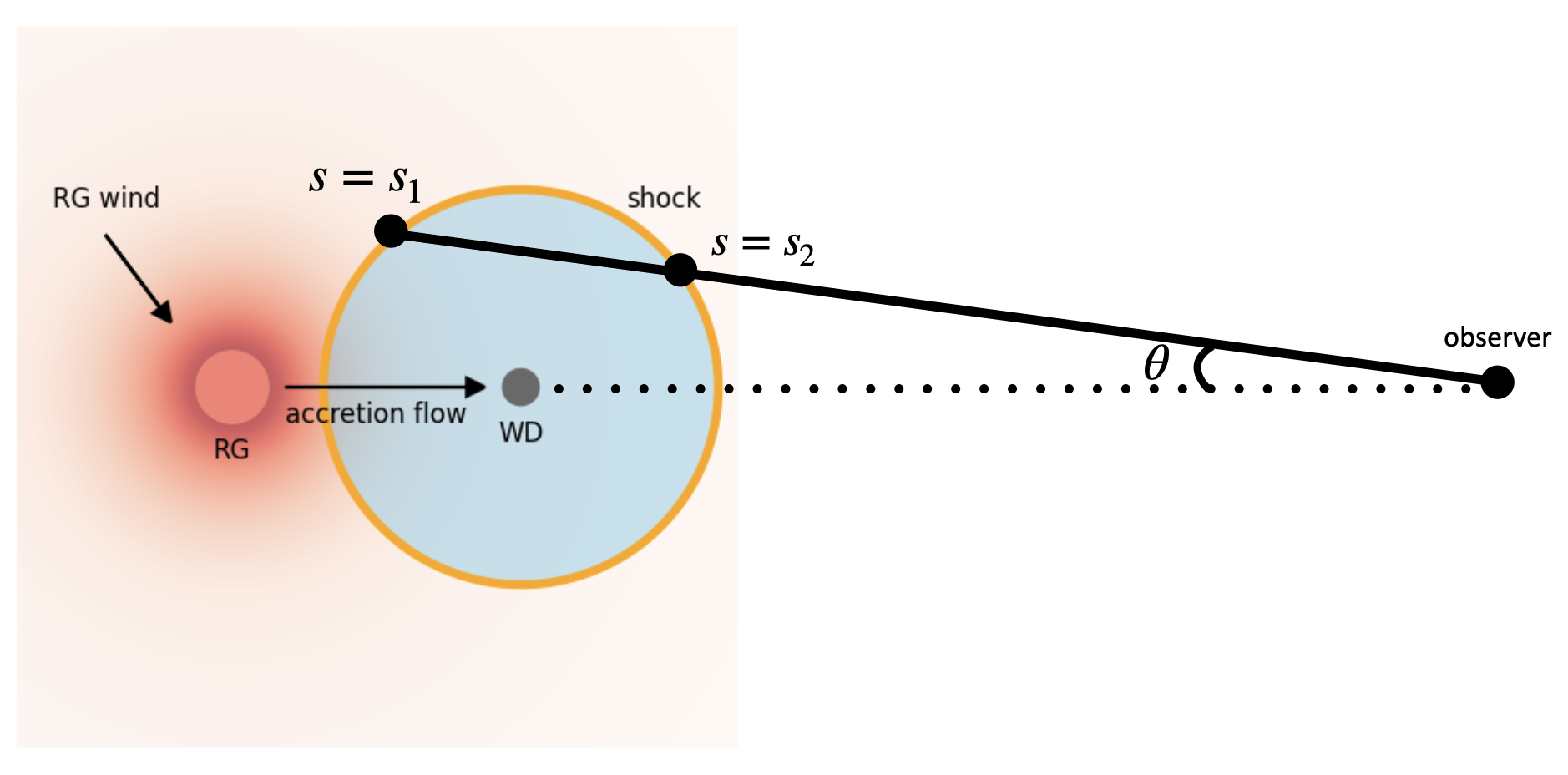}
    \caption{Schematic view of the system. 
    }
    \label{fg:schematic}
\end{figure}

\begin{figure}[h!]
    \centering
	\includegraphics[width=0.48\columnwidth]{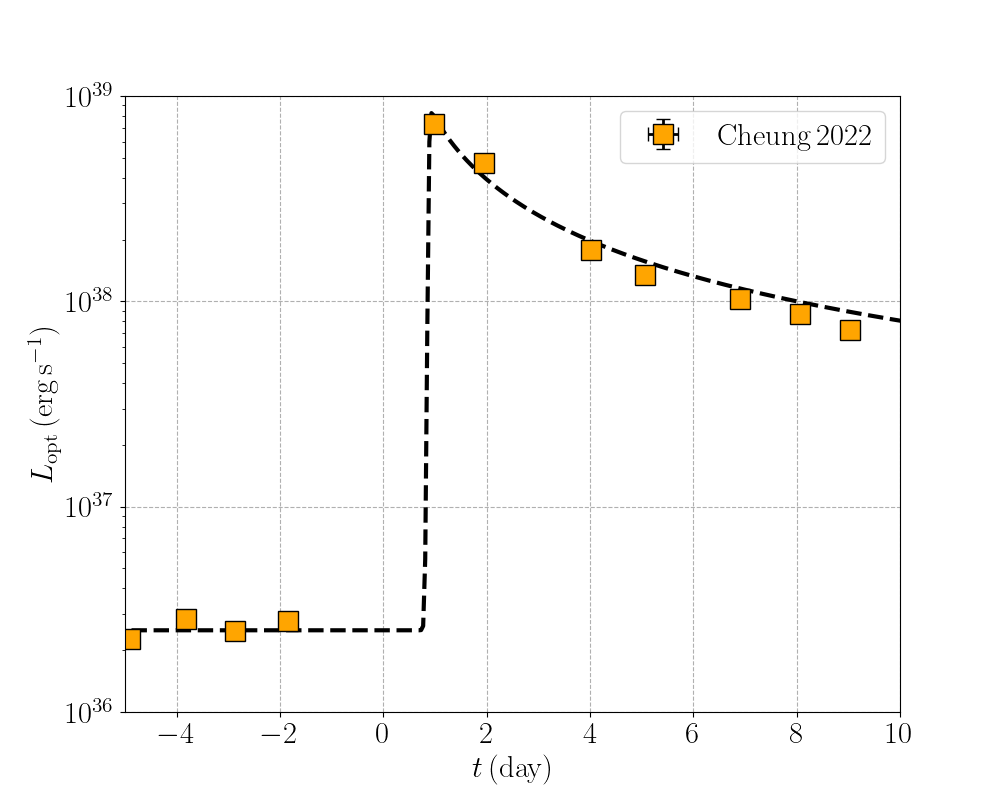}
 	\includegraphics[width=0.48\columnwidth]{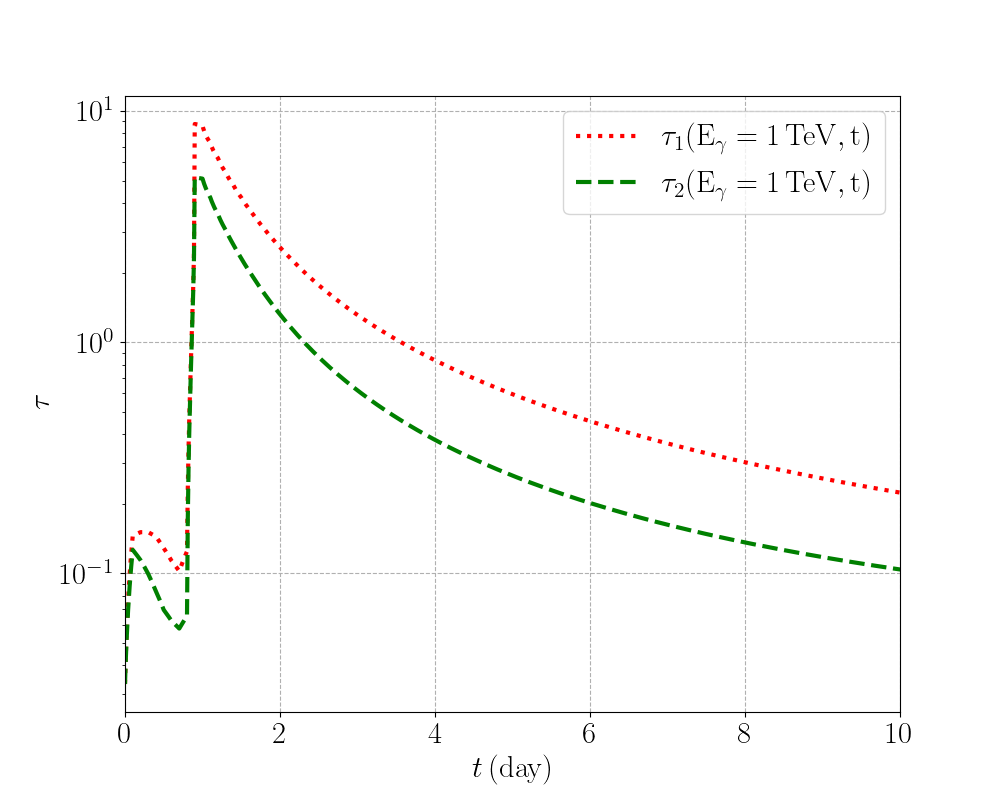}
    \caption{{\it Left}: Luminosity function of optical photons (see Eq. \ref{eq:Lopt}) compared to data collected in \cite{cheung2022} assuming $d_{\rm N}=1.6$ kpc. {\it Right}: Opacities over time estimated from Eq. \ref{eq:tau1} and Eq. \ref{eq:tau2} for $E_\gamma=1$ TeV. 
    }
    \label{fg:luminosity}
\end{figure}

Note that we have introduced also the differential number density of optical photons $f_{\rm opt}(E_{\rm ph}, r, t)$ which, as mentioned above, can be modeled with a black-body distribution as follows:  
\begin{eqnarray}
    f_{\rm opt}(E_{\rm ph},r,t)=u_{\rm opt}(r,t)\frac{15\left(\frac{E_{\rm ph}}{k_B T_{\rm opt}}\right)^2}{ \pi^4(k_B T_{\rm opt})^2\left[\exp\left(\frac{E_{\rm ph}}{k_B T_{\rm opt}}\right)-1\right]},
\end{eqnarray}
where $u(r,t)$ is the optical photon energy density (see Eq. \ref{eq:uOPT}), which can be derived from the optical lightcurve $L_{\rm opt}(t)$ (the optical luminosity as a function of time, see Eq. \ref{eq:Lopt} and Fig. \ref{fg:luminosity}), $k_B$ is the Boltzmann constant, and $T_{\rm opt}$ the black-body temperature.

For each line of sight, we can solve the radiative transfer equation (Eq. \ref{eq:rad_transfer}) to show that $I(E_\gamma,\theta,s\simeq d_{\rm N},t)$ always has two contributions corresponding to gamma rays coming from $s=s_1$ and $s=s_2$ which are the two solutions of $r(\theta,s)=R_{\rm sh}$ (see Fig. \ref{fg:schematic}). We have implicitly chosen the coordinate system such that for all lines of sight $s_1=0$ and $s_2=2\sqrt{R_{\rm sh}-d_{\rm N}^2\sin^2\theta}$. Given this choice of the coordinate system, we can write down the solution of the radiative transfer equation as follows:
\begin{eqnarray}
    I(E_\gamma,\theta,s\simeq d_{\rm N},t)=\frac{R_{\rm sh}I_0(E_\gamma,t)}{\sqrt{R_{\rm sh}^2-d^2\sin^2\theta}}\left[\exp\left(-\int^{d_{\rm N}}_{s_1} \df s \,\eta_{\rm abs}(E_\gamma,r(\theta,s),t)\right)+\exp\left(-\int^{d_{\rm N}}_{s_2} \df s\, \eta_{\rm abs}(E_\gamma,r(\theta,s),t)\right)\right].\n\\
\end{eqnarray}

We can now derive the gamma-ray flux with absorption by integrating the intensity along each line of sight over the solid angle covering the emission region
\begin{eqnarray}    
    \phi(E_\gamma,t)=2\pi\int_0^{R_{\rm sh}/d_{\rm N}}\df \theta \sin\theta\, I(E_\gamma, \theta, s=d_{\rm N}, t).
\end{eqnarray}
The factor $2\pi$ comes from integrating over the azimuthal angle. Note also that we have made the approximation $\arcsin\left(R_{\rm sh}/d_{\rm N}\right)\simeq R_{\rm sh}/d_{\rm N}$ since $d_{\rm N}\gg R_{\rm sh}$. Also, we can show in this way that the gamma-ray flux becomes $\phi_0(E_\gamma,t)$ when the absorption coefficients are set to zero. For ease of discussion, we will express the gamma-ray flux as follows  
\begin{eqnarray}    
    \phi(E_\gamma,t)=\frac{\phi_0(E_\gamma,t)}{2}\left[e^{-\tau_1(E_\gamma,t)}+e^{-\tau_2(E_\gamma,t)}\right],
\end{eqnarray}
where
\begin{eqnarray}
    e^{-\tau_1(E_\gamma,t)}=\left(\frac{d_{\rm N}}{R_{\rm sh}}\right)^2\int_0^{R_{\rm sh}/d_{\rm N}}\df\theta\frac{R_{\rm sh}\sin\theta}{\sqrt{R_{\rm sh}^2-d^2\sin^2\theta}}\exp\left(-\int^{d_{\rm N}}_0 \df s \,\eta_{\rm abs}(E_\gamma,r(\theta,s),t)\right),\label{eq:tau1}
\end{eqnarray}
and
\begin{eqnarray}   
    e^{-\tau_2(E_\gamma,t)}=\left(\frac{d_{\rm N}}{R_{\rm sh}}\right)^2\int_0^{R_{\rm sh}/d_{\rm N}}\df\theta\frac{R_{\rm sh}\sin\theta}{\sqrt{R_{\rm sh}^2-d^2\sin^2\theta}}\exp\left(-\int^{d_{\rm N}}_{2\sqrt{R_{\rm sh}^2-d^2\sin^2\theta}} \df s \,\eta_{\rm abs}(E_\gamma,r(\theta,s),t)\right).\label{eq:tau2}
\end{eqnarray}
We present also estimates of the opacities for $E_\gamma=1$ TeV in Fig. \ref{fg:luminosity}. It is clear from this plot and also from the above equations that gamma rays that have to pass through the shock downstream are more strongly absorbed and, thus, $\tau_1(E_\gamma,t)>\tau_2(E_\gamma,t)$ (see also discussions in Section \ref{sec:rad-transfer}). 

\section{Fit parameters}
\label{appendixB}
\begin{table*}[h!]
\caption{Parameters for fitting the gamma-ray data from the 2021 outburst of RS Oph.}             
\label{table:1}      
\centering          
\begin{tabular}{| c | l | c | c |}     
\hline\hline       
Parameters $\vphantom{^{a}_{a}}$ & Descriptions & Comments & Values\\ 
\hline     
$t_{\rm er}$ & eruption time & \multirow{5}{*}{fixed parameters} & 0 day\\
$r_{\rm orb}$ & orbital radius of RG \citep{brandi2009} & & $1.48$ au\\
$d_{\rm N}$ & distance from RS Oph to Earth \citep{rupen2008} & & $2.45$ kpc\\
$T_{\rm opt}$ & temperature of optical photons \citep{cheung2022} & & $1.1\times 10^4$ K\\
$R_{\rm ph}$ & radius of the photosphere \citep{acciari2022} & & $200\, R_{\odot}$\\
$\dot{M}/v_{\rm wind}$ & mass-loss rate of RG over wind speed of RG \citep{tatischeff2007} & & $2\times 10^{13}$ g/cm\\ 
\hline
$\xi_{\rm p}$ & fraction of shock ram pressure converted into pressure of accelerated protons & \multirow{6}{*}{fitted parameters} & 0.14\\ 
$v_{\rm sh,0}$ & initial speed of the shock & & $3500$ km/s\\
$t_{\rm r}$ & transition time of the shock speed (Eq. \ref{eq:vsh}) & & $2.2$ day\\
$\alpha$ & power-law index of the shock speed (Eq. \ref{eq:vsh}) & & $0.43$\\
$\delta$ & spectral index of the injection spectrum & & 4.2\\
$B_{2,0}$ & magnetic field strength close to RG (Eq. \ref{eq:Bfield}) & & $6.5$ G\\
\hline                  
\end{tabular}
\end{table*}

\bibliography{mybib}{}
\bibliographystyle{aasjournal}

\end{document}